\documentstyle{article}\begin{document}
\title{Statistical physics of the inflaton decaying  in an inhomogeneous random environment}
\author{ Z. Haba\\
Institute of Theoretical Physics, University of Wroclaw,\\ 50-204
Wroclaw, Plac Maxa Borna 9, Poland\\
email:zhab@ift.uni.wroc.pl}\maketitle
\begin{abstract} We derive a stochastic wave equation for an inflaton in an environment
of  an infinite number of fields .  We study solutions of the
linearized stochastic evolution equation in an expanding universe.
The Fokker-Planck equation for the inflaton probability
distribution is derived. The relative entropy (free energy) of the
stochastic wave is defined. The second law of thermodynamics for
the diffusive system is obtained. Gaussian probability
distributions are studied in detail.\end{abstract}

\section{Introduction}
The $\Lambda$CDM model became the standard cosmological model
since the discovery of the universe acceleration
\cite{a1}\cite{a2}. It describes very well the large scale
structure of the universe. The formation of an early universe is
explained by the inflationary models involving scalar fields
(inflatons)\cite{mukhanov}\cite{sas}.  Such models raise the
questions concerning the dynamics  from their early stages till
the present day. The origin of the cosmological constant as a
manifestation of "dark energy" could also be explored.
 In models of the dark sector we hope to explain the "coincidence
problem": why the densities of the dark matter and the dark energy
are of the same order today as well as "the cosmological constant
problem" \cite{wein}: why the cosmological constant is so small.
We assume that the dark matter and dark energy consist of some
unknown particles and fields. They interact in an unknown way with
baryons and the inflaton. The result
 of the
interaction could be seen in a dissipative and diffusive behaviour
of the observed luminous matter. The diffusion approximation does
not depend on the details of the interaction but only on its
strength and "short memory" (Markovian approximation). In this
paper we follow an approach appearing in many papers (see
\cite{peebdm}\cite{peebin}\cite{bolotin}\cite{bertolami}
\cite{shaposh} and references quoted there) describing the dark
matter and the fields responsible for  inflation (inflatons) by
scalar fields. In the $\Lambda$CDM model the universe originates
from the quantum Big Bang. The quantum fluctuations expand forming
the observed galactic structure. The transition to classicality
requires a decoherence. The decoherence can be obtained through an
interaction with an environment. The environment may consist of
any unobservable degrees of freedom. In \cite{hu}\cite{kiefer}
these unobservable variables are the high energy modes of the
fields present in the initial theory. We assume that the
environment consists of an infinite set of scalar fields
interacting with the inflaton . The model is built in close
analogy to the the well-known infinite oscillator model
\cite{ford}\cite{fordkac}\cite{cald}\cite{kleinert} of Brownian
motion. As the model involves an infinite set of unobservable
degrees of freedom the statistical description is unavoidable. We
have an environment of an infinite set of scalar $\chi$ fields
which have an arbitrary initial distributions. We begin their
evolution from an equilibrium thermal state. In such a case we
obtain a random physical system driven by thermal fluctuations of
the environment. We could also take into account  quantum
fluctuations extracting them as high momentum modes of the
inflaton as is done in the Starobinsky stochastic inflation
\cite{starob}. In such a framework a description of a quantum
random evolution is reduced to a classical stochastic process. We
can apply the thermodynamic formalism to the study of evolution of
stochastic systems. In such a framework we can calculate the
probability of a transition from one state to another. In
particular, a vacuum decay can be treated as a stochastic process
leading to  a production of radiation \cite{rad}.

 The plan of this paper is the
following. In sec.2 we derive the stochastic wave equation for an
inflaton interacting with an infinite set of scalar fields in a
homogeneous expanding metric. In sec.3 we briefly discuss a
generalization to inhomogeneous perturbations of the metric
satisfying Einstein-Klein-Gordon equations. In sec.4 we
approximate the non-linear system by a linear inhomogeneous
stochastic wave equation with a space-time dependent mass. The
Fokker-Planck equation for the probablity distribution of the
inflaton is derived. A solution of the Foker-Planck equation in
the form of a Gibbs state with a time-dependent temperature is
obtained. In sec.5 a general linear system is discussed. In sec.6
we obtain partial differential equations for the correlation
functions of this system. In sec.7 Gaussian solutions of the
Fokker-Planck equation and their relation to the equations for
correlation functions are studied. In sec.8 we discuss
thermodynamics of time-dependent (non-equilibrium) diffusive
systems based on the notion of the relative entropy (free energy).
In the Appendix we treat a simple system of stochastic oscillators
in order to show that the formalism works well for this system.
\section{Scalar
fields interacting linearly with an environment} The CMB
observations show that the universe was once in an equilibrium
state. The Hamiltonian dynamics of scalar fields usually discussed
in the model of inflation do not equilibrate. We can achieve an
equilibration if the scalar field interacts with an environment.
We suggest a field theoretic model which is an extension of the
well-known oscillator model discussed in
\cite{ford}\cite{fordkac}\cite{cald}\cite{kleinert}. We consider
the Lagrangian
\begin{equation}\begin{array}{l}
{\cal L}=R\sqrt{g}+\frac{1}{2}\partial_{\mu}\phi\partial^{\mu}\phi
-V(\phi)+\sum_{b}(\frac{1}{2}\partial_{\mu}\chi^{b}\partial^{\mu}\chi^{b}
-\frac{1}{2}m_{b}^{2}\chi^{b}\chi^{b}-\lambda_{b}\phi\chi^{b}-U_{b}(\chi^{b})).
\end{array}\end{equation} Equations of motion for the scalar fields read
\begin{equation}
g^{-\frac{1}{2}}\partial_{\mu}(g^{\frac{1}{2}}\partial^{\mu})\phi=-V^{\prime}-\sum_{b}\lambda_{b}\chi^{b},
\end{equation}
\begin{equation}
g^{-\frac{1}{2}}\partial_{\mu}(g^{\frac{1}{2}}\partial^{\mu})\chi^{b}+m_{b}^{2}\chi^{b}=-\lambda_{b}\phi
-\frac{\partial U_{b}}{\partial \chi^{b}},
\end{equation}where $g_{\mu\nu}$ is the metric tensor and $g=\vert \det[g_{\mu\nu}]\vert$.

We consider the flat expanding metric

\begin{equation}
ds^{2}=dt^{2}-a^{2}d{\bf x}^{2},
\end{equation}
In the metric (4) eq.(3) reads
\begin{equation}
\partial_{t}^{2}\chi^{b}+3H\partial_{t}\chi^{b}-a^{-2}\triangle\chi^{b}+m_{b}^{2}\chi^{b}=-\lambda_{b}\phi
-\frac{\partial U_{b}}{\partial \chi^{b}}\end{equation} We may
choose
\begin{equation}
U_{b}(\chi^{b})=\kappa_{b}(\chi_{b}^{2}-v_{b}^{2})^{2}
\end{equation}
We write
\begin{equation}
\chi=v+a^{-\frac{3}{2}}\tilde{\chi}
\end{equation}

Then\begin{equation}
\partial_{t}^{2}\tilde{\chi}^{b}-a^{-2}\triangle\tilde{\chi}^{b}+\omega_{b}^{2}\tilde{\chi}^{b}=-\lambda_{b}a^{\frac{3}{2}}\phi
+a^{-\frac{3}{2}}o(\tilde{\chi}^{2})\end{equation} where
\begin{equation}
\omega_{b}^{2}=m_{b}^{2}+8\kappa_{b}v_{b}^{2}-\frac{3}{2}\partial_{t}H-\frac{9}{4}H^{2}
\end{equation} We consider large  $a\rightarrow\infty$ so that for a large time
we may  neglect $a^{-2} $ term. Moreover, we assume that
$\omega_{b}^{2}>0$ and that $\omega_{b} $ is approximately
constant (this is exactly so for the de Sitter space and
approximate for power-law expansion when the $H$ dependent term
decays as $t^{-2}$). Then, the solution of eq.(8) is
\begin{equation}
\tilde{\chi}^{b}=\cos(\omega_{b}
t)\tilde{\chi}^{b}_{0}+\sin(\omega_{b}
t)\omega_{b}^{-1}\tilde{\pi}^{b}_{0}-\lambda_{b}
\int_{t_{0}}^{t}\sin(\omega_{b}(
t-s))\omega_{b}^{-1}a(s)^{\frac{3}{2}}\phi_{s}ds
\end{equation}
Inserting the solution of eq.(3) in eq.(2) we obtain an equation
of the form
\begin{equation}\begin{array}{l}
g^{-\frac{1}{2}}\partial_{\mu}(g^{\frac{1}{2}}\partial^{\mu})\phi+V^{\prime}(\phi)
=\int_{t_{0}}^{t}{\cal
K}(t,t^{\prime})\phi(t^{\prime})dt^{\prime}+a^{-\frac{3}{2}}\eta(\chi(0),\tilde{\pi}),
\end{array}\end{equation}where
\begin{equation}
{\cal
K}(t,s)=a(t)^{-\frac{3}{2}}\sum_{b}\lambda_{b}^{2}\sin(\omega_{b}(t-s))\omega_{b}^{-1}a(s)^{\frac{3}{2}}
\end{equation}
and the noise $\eta$ depends linearly on the initial conditions
$(\tilde{\chi}(0),\tilde{\pi}(0))$
\begin{equation}
\eta_{t}=-\sum_{b}\lambda_{b}\cos(\omega_{b}
t)\tilde{\chi}^{b}_{0}+\lambda_{b}\sin(\omega_{b}
t)\omega_{b}^{-1}\tilde{\pi}^{b}_{0}
\end{equation}
If the correlation function of the noise is to be stationary
(depend on time difference) then we need
\begin{equation}
\langle \tilde{\chi}^{b}_{0}\tilde{\chi}^{b}_{0}\rangle=\langle
\omega^{-2}_{b}\tilde{\pi}^{b}_{0}\tilde{\pi}^{b}_{0}\rangle
\end{equation}
Then, (assuming that $\langle \tilde{\chi}\tilde{\pi}\rangle=0$,
true in the classical  Gibbs state) we get
\begin{equation}
\langle \eta_{t}({\bf x})\eta_{s}({\bf
y})\rangle=a^{-3}\sum_{b}\lambda_{b}^{2}\langle
\Big(\omega^{-2}_{b}\cos(\omega_{b}(t-s))\tilde{\pi}^{b}_{0}({\bf
x}),\tilde{\pi}^{b}_{0}({\bf y})\Big)\rangle
\end{equation} We assume a certain probability distribution for initial values. The relation
(14) is  satisfied for classical as well as  quantum Gibbs
distribution with the Hamiltonian
$\tilde{H}_{b}=\frac{1}{2}(\tilde{\pi}^{2}+\omega_{b}^{2}\tilde{\chi}_{b}^{2})$.
In the classical field theory in the Gibbs state  the covariance
of the fields in eq.(14) is
$(-a^{-2}\triangle+\omega_{b}^{2})^{-1}$. If $a^{-2}\triangle=0$
then  this covariance is  approximated by \cite{berera}
$\beta^{-1}\omega_{b}^{-2}\delta({\bf x}-{\bf y})$. We choose
\begin{equation} \lambda_{b}\simeq
\sqrt{\beta}\gamma\pi^{-\frac{1}{2}}\omega_{b}
\end{equation}
Under the assumption (16) and a continuous spectrum of
$\omega_{b}$ in eq.(11) we shall have
\begin{displaymath}\begin{array}{l}
\int_{t_{0}}^{t}ds{\cal
K}(t,s)\phi(s)=-\gamma^{2}\partial_{t}\phi(t)
 -\frac{3}{2}\gamma^{2}H(t)\phi(t) \cr
+\gamma^{2}\delta(0)\phi(t)-\gamma^{2}\delta(t-t_{0})\phi(t_{0})a(t)^{-\frac{3}{2}}a(t_{0})^{\frac{3}{2}}
\end{array}
\end{displaymath}Here, $\delta(t)$ comes from
$\beta^{-1}\sum_{b}\lambda_{b}^{2}\omega_{b}^{-2}\cos(\omega_{b}
t)$; the $\delta(0) $ term is  (an infinite) mass renormalization
which appears already in the Caldeira-Leggett model \cite{cald},
it could be included in $m^{2}$.The last term can be neglected
when $t_{0}$ tends to $-\infty$.
 We shall omit
these terms in further discussion.

In an expanding metric eq.(11) takes the form
\begin{equation}
\partial_{t}^{2}\phi-a^{-2}\triangle\phi+(3H+\gamma^{2})\partial_{t}\phi+
\frac{3}{2}\gamma^{2}H\phi+V^{\prime}(\phi)=\gamma
a^{-\frac{3}{2}}\eta.
 \end{equation}

where
\begin{equation}
\langle \eta(t,{\bf x})\eta(t^{\prime},{\bf
x}^{\prime})\rangle=\delta(t-t^{\prime})K_{t}({\bf x},{\bf
x}^{\prime}).\end{equation} Here, $K ({\bf x},{\bf
x}^{\prime})=\delta({\bf x}-{\bf x}^{\prime})$ comes from the
expectation value of the initial values $\tilde{\chi}$ and
$\tilde{\pi}$ with the neglect of $a^{-2}\triangle$. If we do not
neglect $a^{-2}\triangle$ in eq.(8) then the form of eq.(18) would
be much more complicated (we would obtain a non-local equation).
We make the approximation (18) which preserves the Markov property
and provides stochastic fields which are regular functions of
${\bf x}$ ( it can be considered as a cutoff  ignoring high
momenta components of $\phi$). Eq.(17) has been derived earlier in
\cite{berera}. It is applied in the model of warm inflation
\cite{warm}.
\section{Stochastic equations in a perturbed inhomogeneous metric}
We did not write yet equations for the metric which result from
Lagrangean (1). The power spectrum of inflaton perturbations
depends on the metric
\cite{mukhanov}\cite{sas}\cite{mu}\cite{starpl}\cite{wands} . In
general, the equations for the metric are difficult to solve. They
are solved in perturbation theory. We consider only scalar
perturbations of the homogenous metric using the scalar fields
$A,B,E,\psi$. Then, the metric is expressed in the form
\cite{mukhanov}\cite{wands}
\begin{equation}
ds^{2}==-(1+2A)dt^{2}+2a\partial_{j}Bdx^{j}dt +
a^{2}\Big((1-2\psi)\delta_{ij}+2\partial_{i}\partial_{j}E
\Big)dx^{i}dx^{j}
\end{equation}
Inserting the metric in eqs.(2)-(3) we obtain
\begin{equation}
\partial_{t}^{2}\phi+(3H+\Gamma)\partial_{t}\phi-a^{-2}\triangle\phi
+V^{\prime}(\phi)(1+2A)=-\sum_{b}\lambda_{b}\chi_{b}(1+2A)
\end{equation}
and
\begin{equation}
\partial_{t}^{2}\chi_{b}+(3H+\Gamma)\partial_{t}\chi_{b}-a^{-2}\triangle\chi_{b}
+m_{b}^{2}\chi_{b}(1+2A)=-\lambda_{b}\phi(1+2A)-\frac{\partial
U_{b}}{\partial \chi_{b}}(1+2A)
\end{equation}
where
\begin{equation}
-\Gamma=\partial_{t}A+3\partial_{t}\psi-a^{-2}\triangle(a^{2}\partial_{t}E-aB)
\end{equation}
Let
\begin{equation}
\chi_{b}=\exp(\sigma)\tilde{\chi}_{b}+v_{b}
\end{equation}
with
\begin{equation}
\partial_{t}\sigma=-\frac{1}{2}(3H+\Gamma)
\end{equation}
Then, in the approximation neglecting higher powers of
$\tilde{\chi}$ in $U$
\begin{equation}
\partial_{t}^{2}\tilde{\chi}_{b}-a^{-2}\triangle\tilde{\chi}_{b}
+\Omega_{b}^{2}\chi_{b}=-\lambda_{b}\exp(-\sigma)(1+2A)\phi
\end{equation}
where
\begin{equation}
\Omega_{b}^{2}=(m_{b}^{2}+8v_{b}^{2}\kappa_{b})(1+2A)-\frac{1}{4}(3H+\Gamma)^{2}-\frac{1}{2}\partial_{t}(3H+\Gamma)
\end{equation}
We again assume that $\Omega^{2}\simeq
m_{b}^{2}+8v_{b}^{2}\kappa_{b}$. Then, in the derivation of
eq.(17) for $\phi$ the only change comes from the differentiation
of $\exp(\sigma)$ inside the integral (11) and the factor $(1+2A)$
multiplying the fields. so,
\begin{equation}
\frac{3}{2}\gamma^{2}H\phi\rightarrow\frac{1}{2}\gamma^{2}(3H+\Gamma)\phi(1+2A)^{2}
\end{equation}and
\begin{equation}
\gamma^{2}\partial_{t}\phi\rightarrow
\gamma^{2}(1+2A)\partial_{t}((1+2A)\phi)
\end{equation}

 Hence, our final equation is (for a general theory of a
 stochastic wave equation on a Riemannian manifold see
 \cite{brzezniak} and references cited there)
\begin{equation}\begin{array}{l}
\partial_{t}^{2}\phi+(3H+\gamma^{2}+\Gamma)\partial_{t}\phi-a^{-2}\triangle\phi+\frac{1}{2}\gamma^{2}(3H+\Gamma)\phi
\cr
+6\gamma^{2}AH\phi+2\gamma^{2}\partial_{t}A\phi+4\gamma^{2}A\partial_{t}\phi+V^{\prime}(\phi)(1+2A)=\gamma
(1+2A)a^{-\frac{3}{2}}\eta \end{array}\end{equation}
\section{A simplified system of a decaying inflaton}
The metric $(A,B, \psi,E)$ can be expressed by $\phi$ from
Einstein equations resulting from the Lagragean (1). We expand
inflaton equation with a potential $V$ around the classical
solution of Klein-Gordon-Einstein equation. The linearized version
of the equation for fluctuations takes the form of the
Klein-Gordon equation with a space-time dependent mass
\cite{mukhanov}\cite{sas}\cite{hwang}\cite{wands}\cite{pot}
\begin{displaymath}
\partial_{t}\phi=\Pi
\end{displaymath}\begin{equation}
 d\Pi+(3H+\gamma^{2})\Pi dt+\frac{3}{2}\gamma^{2}H\phi dt+\nu\phi dt +a^{-2}\triangle\phi dt=\gamma a^{-\frac{3}{2}}dB
\end{equation}where we write $\eta=\frac{dB}{dt}$ and treat (30)
as Ito stochastic differential equation \cite{ikeda}. The function
$\nu$ depends on the potential $V$ in eq.(17) and on the choice of
coordinates $(t,x)$ (the choice of gauge \cite{bar}). We do not
discuss $\nu$ in this paper. We consider in this section the
simplified version of eq.(30) without the $\phi$ terms

\begin{equation}
 d\Pi+(3H+\gamma^{2})\Pi dt=\gamma a^{-\frac{3}{2}}dB
\end{equation}

We define the energy density
\begin{equation}
\rho=\frac{1}{2}\Pi^{2}
\end{equation}
Then, from eq.(31) applying the stochastic calculus
\cite{ikeda}\cite{simon}and eq.(31) we obtain
\begin{displaymath}
d\Pi^{2}=2\Pi d\Pi+d\Pi d\Pi=-2(3H+\gamma^{2})\Pi^{2}dt+ 2\gamma
a^{-\frac{-3}{2}}\Pi dB+\gamma^{2}a^{-3}K(x,x)dt
\end{displaymath}
We may first integrate this equation and use $\langle
\int_{0}^{t}fdB\rangle=0$ for the Ito integral. Differentiating
the expectation value over $t$ we obtain

\begin{equation} d
\langle\rho\rangle+6H \langle\rho\rangle dt=-
2\gamma^{2}\langle\rho\rangle dt +\frac{1}{2}\gamma^{2}K(x,x)a^{-3} dt
\end{equation}
Eq.(33) describes the inflaton density  with $w=\frac{\rho}{p}=1$
and a cosmological term varying with the speed $a^{-3}$. The term
$- 2\gamma^{2}\langle\rho\rangle $ violates the energy
conservation of the inflaton . It describes a decay of the
inflaton into the $\chi$ fields. If we couple the $\chi$ fields to
radiation then if $\chi$ fields are invisible the observable
effect will be detected as  a production of radiation from the
decay of the inflaton \cite{rad}\cite{fang}.

For the stochastic system (31) the Fokker-Planck equation reads
\begin{equation}\begin{array}{l}
\partial_{t}P=\frac{\gamma^{2}}{2}\int d{\bf
x} d{\bf x}^{\prime}{\cal G}_{t}({\bf x},{\bf
x}^{\prime})\frac{\delta^{2}}{\delta\Pi({\bf x})\delta\Pi({\bf
x}^{\prime})} P+\int d{\bf x}\frac{\delta}{\delta\Pi({\bf
x})}(3H+\gamma^{2})\Pi P
 .\end{array}\end{equation}Then, the Gaussian solution is
\begin{equation}
P=L\exp(-\frac{1}{2}\int \sqrt{g} \Pi\beta\Pi)
\end{equation}where $\sqrt{g}=a^{3}$.
It can be checked that\begin{equation}
\beta=\exp(2\gamma^{2}t)a^{3}\Big(R+\gamma^{2}\int_{0}^{t}dsa(s)^{3}\exp(2\gamma^{2}s)\Big)^{-1}
\end{equation} and \begin{equation}
L^{-1}\partial_{t}L=-\frac{1}{2}\gamma^{2}a^{-3}Tr(K\beta)+(3H+\gamma^{2})\delta(0)
\int d{\bf x}.
\end{equation} This normalization factor is infinite (needs
renormalization) but the value of $L$ does not appear in the
expectation values  $\langle F\rangle=(\int P)^{-1}\int P F$.

$\beta$ has the meaning of the inverse temperature. The dependence
(36) of the temperature of the diffusing system on the scale
factor $a$ has been derived (for $\gamma=0$ and arbitrary $w$) in
 \cite{habadiff}\cite{hss}\cite{habagrg}for any system with $w=1$ and the
$a^{-3}$ correction  (33) to the cosmological term (in eq.(36)
$w=1$)(for time dependent cosmological term see
\cite{review}\cite{weinberg}\cite{ss}) .

\section{The linearized wave equation}We can rewrite eqs.(4)-(5) in a way that they do not contain first
order time derivatives of fields. Let \begin{displaymath}
\phi=a^{-\frac{3}{2}}\exp(-\frac{1}{2}\gamma^{2}t)\Phi\end{displaymath}
Then, the linearized version of the inflaton equation expanded
around the classical solution (with an account of  metric  perturbations ) reads
\begin{displaymath}
\partial_{t}\Phi=\Pi
\end{displaymath}
\begin{equation}\begin{array}{l}
\partial_{t}\Pi+K^{2}\Phi-\frac{3}{2}\partial_{t}H\Phi
-\frac{9}{4}H^{2}\Phi-\frac{1}{4}\gamma^{2}\Phi+\nu\Phi\cr=
\partial_{t}\Pi+K^{2}\Phi+\tilde{\nu}\Phi =\gamma
a^{\frac{3}{2}}\exp(\frac{1}{2}\gamma^{2}t)\eta
\end{array}\end{equation}where
\begin{equation}
K^{2}=-a^{-2}\triangle +m^{2}
\end{equation}
 For the stochastic system
(38) the Fokker-Planck equation reads
\begin{equation}\begin{array}{l}
\partial_{t}P=\frac{\gamma^{2}}{2}\int d{\bf
x} d{\bf x}^{\prime}{\cal G}_{t}({\bf x},{\bf
x}^{\prime})\frac{\delta^{2}}{\delta\Pi({\bf x})\delta\Pi({\bf
x}^{\prime})}P\cr+\int d{\bf x}(K^{2}\Phi+\tilde{\nu}
\Phi)\frac{\delta}{\delta\Pi({\bf x})}P
 -\int d{\bf x}\Pi({\bf
x})\frac{\delta}{\delta\Phi({\bf x})}P\equiv {\cal
A}P.\end{array}\end{equation} where
\begin{equation}
{\cal G}_{t}({\bf x},{\bf
x}^{\prime})=a^{3}\exp(\gamma^{2}t)K_{t}({\bf x},{\bf x}^{\prime})
\end{equation}
and $\nu$ depends on the classical solution in the potential $V$
\cite{hwang}\cite{wands}.

 We may write the noise  in the Fourier momentum space
\begin{equation}
\langle \eta(t,{\bf k})\eta(t^{\prime},{\bf
k}^{\prime})\rangle={\cal G}_{t}({\bf k})\delta({\bf k}+{\bf
k}^{\prime})\delta(t-t^{\prime})
\end{equation}
Then, eq.(38) is rewritten as an ordinary (instead of partial)
differential equation.

\section{A differential equation for correlations}
Let us consider eq.(38) expressed in the form
\begin{equation}
\partial_{t}\Phi=\Pi
\end{equation}
\begin{equation}\partial_{t}\Pi=A\Phi+\gamma a^{\frac{3}{2}}\exp(\frac{1}{2}\gamma^{2}t)\eta
\end{equation}where
\begin{equation}
-A=K^{2}+\tilde{\nu}= -a^{-2}\triangle + m^{2}+\tilde{\nu}
\end{equation}
Let us denote
\begin{equation}
\langle\Phi_{t}({\bf x})\Phi_{t}({\bf y})\rangle=C_{t}({\bf
x},{\bf y})
\end{equation}\begin{equation}
\langle\Phi_{t}({\bf x})\Pi_{t}({\bf y})\rangle=E_{t}({\bf x},{\bf
y})
\end{equation}\begin{equation}
\langle\Pi_{t}({\bf x})\Pi_{t}({\bf y})\rangle=D_{t}({\bf x},{\bf
y})
\end{equation}
Using the stochastic calculus \cite{ikeda}\cite{simon} and taking
the expectation value we get a system of differential equations
for the correlation functions
 \begin{equation}\partial_{t}C_{t}({\bf x},{\bf y})=E_{t}({\bf x},{\bf y})
 +E_{t}({\bf y},{\bf x})\end{equation}
 \begin{equation}\partial_{t}D_{t}({\bf x},{\bf y})=A_{x}E_{t}({\bf x},{\bf y})+
 A_{y}E_{t}({\bf y},{\bf x})+\gamma^{2}{\cal G}_{t}({\bf x},{\bf y})\end{equation}
 \begin{equation}
 \partial_{t}E_{t}({\bf y},{\bf x})=A_{x}C_{t}({\bf x},{\bf y})+D_{t}({\bf x},{\bf y})\end{equation}
If the system is translation invariant
 then we can Fourier transform these equations
obtaining a system of ordinary differential equations for Fourier
transforms
\begin{equation}
\partial_{t}C_{t}(k)=E_{t}(k)+E_{t}(-k)
\end{equation}
\begin{equation}
\partial_{t}D_{t}(k)=-(a^{-2}k^{2}+m^{2}+\tilde{\nu})(E_{t}(k)+E_{t}(-k))+\gamma^{2}{\cal G}_{t}(k)
\end{equation}
\begin{equation}
\partial_{t}E_{t}(k)=-(a^{-2}k^{2}+m^{2}+\tilde{\nu})C_{t}(k)+D_{t}(k)
\end{equation}
where ${\cal G}_{t}(k)$ is defined in eq.(42).
\section{Gaussian solutions of the
Fokker-Planck equation} We look for a solution of the
Fokker-Planck equation (40) in the form
\begin{equation}\begin{array}{l}
P_{t}^{I}=L(t)\exp\Big(-\gamma^{-2}\int d{\bf x}d{\bf
x}^{\prime}\Big(\frac{1}{2}\Pi\beta_{1}(t,{\bf x},{\bf
x}^{\prime})\Pi+\Pi\beta_{2}(t,{\bf x},{\bf
x}^{\prime})\Phi+\frac{1}{2}\Phi\beta_{3}(t,{\bf x},{\bf
x}^{\prime})\Phi\Big)\cr+\int d{\bf x}M\Phi+\int d{\bf
x}N\Pi\Big).
\end{array}\end{equation}or in the momentum space
\begin{equation}\begin{array}{l}
P_{t}^{I}=L(t)\exp\Big(-\gamma^{-2}\int d{\bf
k}\Big(\frac{1}{2}\Pi\beta_{1}(t,{\bf k})\Pi+\Pi\beta_{2}(t,{\bf
k})\Phi+\frac{1}{2} \Phi\beta_{3}(t,{\bf k})\Phi\Big)\cr+\int
d{\bf k }M({\bf k})\Phi(-{\bf k})+ \int d{\bf k}N({\bf
k})\Pi(-{\bf k})\Big). \end{array}\end{equation} In the
configuration space $\beta$ is an operator and in the momentum
space a function of ${\bf k}$. $L(t)$ is determined by
normalization or directly from the Fokker-Planck equation
\begin{equation}
L^{-1}\partial_{t}L=-\frac{1}{2}\int d{\bf x}d{\bf
x}^{\prime}{\cal G}_{t}({\bf x},{\bf x}^{\prime})\beta_{1}({\bf
x},{\bf x}^{\prime}) \end{equation}
The equations for $\beta$ read

\begin{equation}
\partial_{t}\beta_{1}=-\beta_{1}{\cal G}_{t}\beta_{1}-2\beta_{2}
\end{equation}
\begin{equation}
\partial_{t}\beta_{2}=-\beta_{2}{\cal
G}_{t}\beta_{1}+\omega^{2}\beta_{1}-\beta_{3}
\end{equation}
\begin{equation}
\partial_{t}\beta_{3}=-\beta_{2}{\cal G}_{t}\beta_{2}+2\omega^{2}\beta_{2}
\end{equation}where
\begin{equation}
\omega^{2}=a^{-2}k^{2}+m^{2}+\tilde{\nu}
\end{equation}
We skip the equations for $M$ and $N$.

 It is useful to
introduce instead of $\beta_{j}$ the variables (defined by Fourier
transforms of $\beta$)
\begin{equation}
X=\beta_{3}-\frac{\beta_{2}^{2}}{\beta_{1}}
\end{equation}\begin{equation}
Y=\frac{\beta_{2}}{\beta_{1}}
\end{equation}\begin{equation}
Z=\frac{\beta_{3}}{\beta_{1}}
\end{equation}
We can invert these relations
\begin{equation}
\beta_{1}=\frac{X}{Z-Y^{2}}
\end{equation}\begin{equation}
\beta_{2}=\frac{XY}{Z-Y^{2}}
\end{equation}\begin{equation}
\beta_{3}=\frac{ZX}{Z-Y^{2}}
\end{equation}

 $P_{t}^{I}$ can be expressed as
\begin{equation}
P_{t}^{I}=L(t)\exp\Big(-\int d{\bf x} d{\bf x}^{\prime}
\frac{1}{2}\gamma^{-2}\Big((\Pi+Y\Phi)\beta_{1}(\Pi+Y\Phi)+\Phi
X\Phi)\Big)\Big).
\end{equation}
From eq.(68) it can be seen that the probability distribution is
diagonal in the variables $\Phi$ and $\Pi+Y\Phi$. So, we obtain
the expectation value $\langle (\Pi+Y\Phi)({\bf
x})(\Pi+Y\Phi)({\bf
x}^{\prime})\rangle=\gamma^{2}\beta_{1}^{-1}({\bf x},{\bf
x}^{\prime})$.

 Assume that we calculate the
expectation values at time $t$
\begin{equation}
D=\langle \Pi^{2}_{t}\rangle=\gamma^{2}\beta_{3}(
\beta_{1}\beta_{3}-\beta_{2}^{2})^{-1}=\gamma^{2}ZX^{-1}
\end{equation}
\begin{equation}
C=\langle \Phi^{2}_{t}\rangle=\gamma^{2}\beta_{1}(
\beta_{1}\beta_{3}-\beta_{2}^{2})^{-1}=\gamma^{2}X^{-1}
\end{equation}
\begin{equation}
E=\langle \Phi_{t}\Pi_{t}\rangle=-\gamma^{2}\beta_{2}(
\beta_{1}\beta_{3}-\beta_{2}^{2})^{-1}=-\gamma^{2}YX^{-1}
\end{equation}
$X,Y,Z$ can be expressed by $D,C,E$ as
\begin{equation}
Z=DC^{-1}
\end{equation}
\begin{equation}
Y=-EC^{-1}
\end{equation}
\begin{equation}
X=\gamma^{2}C^{-1}
\end{equation}

If we know $D,E,C$  then we can express
\begin{equation}
\beta_{1}=C(DC-E^{2})^{-1}
\end{equation}
\begin{equation}\beta_{2}=-E(DC-E^{2})^{-1}
\end{equation}
\begin{equation}
\beta_{3}=D(DC-E^{2})^{-1}
\end{equation}
Note that $\beta_{2}(t=0)=0$ means $E(t=0)=0$.

The relations (63)-(77) allow to relate the solutions of the
stochastic equation (38) with the solutions of the differential
equations (58)-(60) and the solution of the Fokker-Planck equation
(40). In fact, the solution $P^{I}$ can be expressed by the
Fokker-Planck transition function $P_{t}$ (which is defined by the
solution of the stochastic equation (38)\cite{ikeda}) as follows
\begin{equation}
P^{I}_{t}(\phi,\Pi)=\int
d\phi^{\prime}d\Pi^{\prime}P^{I}_{0}(\phi^{\prime},\Pi^{\prime})P_{t}(\phi^{\prime},\Pi^{\prime};\phi,\Pi)
\end{equation}
The expectation values of the solution of the stochastic equation
with the initial condition $(\phi^{\prime},\Pi^{\prime})$ is
\begin{displaymath}
\Big\langle
F\Big(\phi_{t}(\phi^{\prime},\Pi^{\prime}),\Pi_{t}(\phi^{\prime},\Pi^{\prime})\Big)\Big\rangle=
\int d\phi d\Pi
P_{t}(\phi^{\prime},\Pi^{\prime};\phi,\Pi)F(\phi,\Pi)
\end{displaymath}
Hence,
\begin{equation}\begin{array}{l}
\int
d\phi^{\prime}d\Pi^{\prime}P^{I}_{0}(\phi^{\prime},\Pi^{\prime})\langle
F(\phi_{t}(\phi^{\prime},\Pi^{\prime}),\Pi_{t}(\phi^{\prime},\Pi^{\prime})\rangle\cr=\int
d\phi^{\prime}d\Pi^{\prime} P^{I}_{0}(\phi^{\prime},\Pi^{\prime})
\int d\phi d\Pi
P_{t}(\phi^{\prime},\Pi^{\prime};\phi,\Pi)F(\phi,\Pi)=\int d\phi
d\Pi P_{t}^{I}(\phi,\Pi)F(\phi,\Pi)\end{array}
\end{equation}
Note that the initial value $P_{0}^{I}$ in eq.(79) according to
eq.(55) is determined by the initial values of $\beta_{j}$. A
possible choice for the initial value is the thermal Gibbs
distribution
\begin{displaymath}
P_{0}^{I}=\exp\Big(-\frac{1}{2T}\int d{\bf
x}(\Pi^{2}+(\nabla\phi)^{2}+m^{2}\phi^{2})\Big)
\end{displaymath}
which corresponds to the initial condition $\beta_{2}(t=0)=0$,
$\beta_{1}(t=0,{\bf x},{\bf x}^{\prime})=\frac{1}{T}\delta({\bf
x}-{\bf x}^{\prime})$ and $\beta_{3}(t=0,{\bf x},{\bf
x}^{\prime})=\frac{1}{T}\triangle\delta({\bf x}-{\bf
x}^{\prime})$. We could also consider the initial probability
distribution
\begin{displaymath}
P_{0}^{I}=\exp\Big(-\frac{1}{2\sigma^{2}}\int d{\bf x}(\phi({\bf
x})-v)^{2}\Big)\end{displaymath} describing the field concentrated
at $v$. Then, in eq.(55) $M(t=0,{\bf x}))=\sigma^{-2}v$. In such a
case, the probability distribution (55) describes the probability
of the transition from $v$ to $\phi$ (see \cite{guth} for such
calculations in quantum mechanics).

\section{The relative entropy} Assume we have a functional
equation of the form (like eq.(40))
\begin{equation}
\begin{array}{l}
\partial_{t}P=\frac{1}{2}\int d{\bf
x} d{\bf x}^{\prime}{\cal D}_{t}({\bf x},{\bf
x}^{\prime})\frac{\delta^{2}}{\delta\Pi({\bf x})\delta\Pi({\bf
x}^{\prime})}P\cr+\int d{\bf x} \frac{\delta}{\delta\Pi({\bf
x})}D_{1}(\phi({\bf x}),\Pi({\bf x}))P +\int d{\bf
x}\frac{\delta}{\delta\Phi({\bf x})}D_{2}(\Phi({\bf x}),\Pi({\bf
x}))P\equiv {\cal A}P \end{array}\end{equation}  Let us assume
that we have two solutions $P_{1}$ and $P_{2}$ of this equation.
Define the relative entropy
\begin{equation}
F=\int d\Phi d\Pi
Z_{1}^{-1}P_{1}\ln\Big(Z_{2}Z_{1}^{-1}P_{1}P_{2}^{-1}\Big)
\end{equation}
where
\begin{equation}
Z_{1}=\int d\Phi d\Pi P_{1}
\end{equation}
\begin{equation}
Z_{2}=\int d\Phi d\Pi P_{2}\end{equation} From the definition of
$F$ it follows that \cite{risken}
\begin{equation}
F\geq 0
\end{equation} Calculation of the
time derivative  of $F$ gives
\begin{equation}
\begin{array}{l}
\partial_{t}F=-\frac{1}{2}\int d\phi d\Pi  P_{1}\Big(P_{2}P_{1}^{-1}\Big)^{2}d{\bf
x}d{\bf x}^{\prime}{\cal D}_{t}({\bf x},{\bf
x}^{\prime})\frac{\delta}{\delta\Pi({\bf
x}))}(P_{1}P_{2}^{-1})\frac{\delta}{\delta\Pi({\bf
x}^{\prime})}(P_{1}P_{2}^{-1})\leq 0
\end{array}\end{equation}
We choose $P_{2}=P^{I}$ (then $Z_{2}=1$ because $L(t)$ is the
normalization factor). Then, we define the entropy (the entropy of
inflaton and gravitational perturbations has been discussed
earlier in \cite{bra1}-\cite{bra2})
\begin{equation}
S=-Z^{-1}\int d\Phi d\Pi P\ln(Z^{-1}P)
\end{equation}
 Using eq.(80) we calculate the time derivative\begin{equation}
\partial_{t}S=-Z^{-1}\int d\Phi d\Pi {\cal A}P\ln P-
Z^{-1}\int d\Phi d\Pi {\cal A} P
\end{equation}
The second term is zero, whereas the first term is  equal to
\begin{equation}
\begin{array}{l}
\partial_{t}S=\frac{1}{2}\int d\Phi d\Pi  P^{-1}d{\bf
x}d{\bf x}^{\prime}{\cal D}_{t}({\bf x},{\bf
x}^{\prime})\frac{\delta}{\delta\Pi({\bf
x})}P\frac{\delta}{\delta\Pi({\bf x}^{\prime})}P\cr+\int d\phi
d\Pi \int d{\bf x}\Big(D_{1}\frac{\delta P}{\delta \Pi({\bf
x})}+D_{2}\frac{\delta P}{\delta \Phi({\bf
x}})\Big)\end{array}\end{equation} In eq.(88) the first term is
positive whereas the second term depends on the dynamics (it is
vanishing for Hamiltonian dynamics). Using the formula (55) for
$P_{2}=P^{I}$ we obtain
\begin{equation}
 F+S=Z^{-1}\int d\Phi d\Pi
 P\Big(\frac{1}{2}\gamma^{-2}\Pi\beta_{1}\Pi+\gamma^{-2}\Phi\beta_{2}\Pi+\frac{1}{2}\gamma^{-2}\Phi\beta_{3}\Phi\Big)
 -\ln L(t)
\end{equation}
The formula (89) has a thermodynamic meaning relating the sum of
free energy $F$ and  entropy $S$ to the internal energy expressed
by the rhs of eq.(89). At the initial time (with the initial
conditions discussed at the end of sec.7) the rhs of eq.(89) is
the mean value of the energy
\begin{displaymath}
U_{0}=\frac{1}{2}\int d{\bf x}(\Pi^{2}+(\nabla\Phi)^{2}+m^{2}\Phi^{2})
\end{displaymath}
In the static universe we would have an equilibrium distribution
as $P_{2}$. In such a case the thermodynamic relation (88) would
describe the standard version of the second law of thermodynamics
of diffusing systems. $F$ with $\partial_{t}F\leq 0$ in eq.(85)
would show the approach to equilibrium. In the expanding universe
the relation (89) can serve for a comparison of various
probability measures starting from different initial conditions.
\section{Summary}
The main source of observational data \cite{a1}-\cite{a2} comes
from measurements of the cosmic microwave background (CMB) and
observations of galaxies evolution (including galaxies
distribution).
 The  CMB spectrum and its fluctuations are the test ground for
 models involving quantum and thermal fluctuations.
A simplified description of an interaction
 of a relativistic system with an environment leads to a stochastic wave equation for an inflaton
 generating the expansion (inflation) of the universe.
 We considered a linearization of the wave equation.
 We discussed the Fokker-Planck equation for the probability distribution of the inflaton.
 Gaussian solutions of the Fokker-Planck equation for linearized systems can be treated
 as Gibbs states with a time-dependent temperature.
 The model leads to a formula for density and temperature evolution.
 We have derived the density evolution law in eq.(33) and
 the temperature evolution in eq.(36) in a simplified model. In order to
 obtain the results in the complete model we would have to solve (numerically) equations of sec.7.
The comparison of density evolution (36) with observations is
discussed in \cite{hss} and in similar models with the decaying
vacuum ( see \cite{review}\cite{ss} and references cited there).
The model allows to calculate (and compare with observations) the
power spectrum resulting from thermal fluctuations which may go
beyond the approximations applied in the warm inflation of
ref.\cite{warm}.
 We have
 introduced a thermodynamic description of the expanding diffusive
 systems
in terms of the relative entropy (free energy) and entropy. The
state of a stochastic system can be identified with its
probability distribution. The relative entropy allows to compare
the evolution of the probability distributions with different
initial conditions. In this sense relative entropy can be treated
as a quantitative measure of a decay of one state into another
state (as an alternative to a quantum description of vacuum decay
in cosmology \cite{dent}\cite{urban}).

{\bf Acknowledgements}

Interesting discussions with Zdzislaw Brzezniak on  stochastic
wave equations  during my stay at York University are gratefully
acknowledged
  \section{Appendix:Statistical physics of a static finite
dimensional model} A finite dimensional analog of the wave
equation is (${\bf x}\in R^{n}$)
\begin{displaymath}
\frac{dx^{k}}{dt}=p^{k}
\end{displaymath}
\begin{equation}
\frac{dp^{k}}{dt}=-\Gamma p^{k}-\omega^{2}x^{k}+\gamma \eta^{k}
\end{equation}
The Fokker-Planck equation reads
\begin{equation}
\partial_{t}P=-p^{k}\frac{\partial }{\partial x^{k}}+\frac{\partial }{\partial
p^{k}}(\Gamma
p^{k}+\omega^{2}x^{k})P+\frac{\gamma^{2}}{2}\frac{\partial^{2}P}{\partial
p^{k}\partial p^{k}}
\end{equation}
The stationary solution is
\begin{equation}
P_{\infty}=\exp\Big(-\frac{\Gamma}{\gamma^{2}}({\bf
p}^{2}+\omega^{2}{\bf x}^{2})\Big)=\exp(-\frac{{\cal E}}{T})
\end{equation}where ${\cal E}$ is the energy of the oscillator.
It describes a Gibbs state with the temperature
\begin{equation}
T=\frac{\gamma^{2}}{2\Gamma}
\end{equation}
We look for a solution of eq.(91) in the form
\begin{equation}
P_{t}=L(t)\exp\Big( -\frac{1}{2}\alpha_{1}{\bf
p}^{2}-\alpha_{2}{\bf x}{\bf p}-\frac{1}{2}\alpha_{3}{\bf x}^{2}
\Big)
\end{equation}
Then
\begin{equation}
L^{-1}\partial_{t}L=-\frac{n}{2}\gamma^{2}\alpha_{1}+\Gamma n
\end{equation}
\begin{equation}
\partial_{t}\alpha_{1}=-2\alpha_{2}-\gamma^{2}\alpha_{1}^{2}+2\Gamma\alpha_{1}
\end{equation}\begin{equation}
\partial_{t}\alpha_{2}=-\alpha_{3}-\gamma^{2}\alpha_{1}\alpha_{2}+\Gamma\alpha_{2}+\omega^{2}\alpha_{1}
\end{equation}\begin{equation}
\partial_{t}\alpha_{3}=-\gamma^{2}\alpha_{2}^{2}+2\omega^{2}\alpha_{2}
\end{equation}
Let us write
\begin{equation}
x^{k}=\exp(-\frac{\Gamma}{2}t)y^{k}
\end{equation}
Then, the stochastic equation for $y$ reads
\begin{equation}
\frac{d^{2}y^{k}}{dt^{2}}=-\Omega^{2}y^{k}+\gamma\exp(\frac{\Gamma}{2}t)\eta^{k}
\end{equation}
where
\begin{equation}
\Omega^{2}=\omega^{2}-\frac{\Gamma^{2}}{4}
\end{equation}
The solution of eq.(100) is
\begin{equation}
y^{k}(t)=\cos(\Omega t)y^{k}(0)+\sin(\Omega
t)\Omega^{-1}\partial_{t}y^{k}(0)+\gamma
\int_{0}^{t}\sin(\Omega(t-s))\Omega^{-1}\exp(\frac{\Gamma}{2}s)w(s)ds
\end{equation}
We can easily calculate the correlation functions of $x^{k}(s)$
and $p^{k}(s)$ in two ways: either from the stochastic equations
or using the probability distribution $P_{t}$ resulting from the
solution of the differential equations (96)-(98).

\end{document}